\documentclass[preprint,showpacs,preprintnumbers,amssymb,superscriptaddress,aps,prd,nofootinbib,11pt]{revtex4-1}

\selectfont 

\usepackage{graphicx}       
\usepackage{dcolumn}        
\usepackage{bm}             
\usepackage{psfrag}
\usepackage{tensor}
\usepackage[usenames]{color}
\definecolor{navyblue}{rgb}{0.0, 0.0, 0.5}
\usepackage[linktocpage,colorlinks=true,allcolors=navyblue]{hyperref}
\usepackage{amsmath}

\newcommand{\beq}{\begin{equation}}
\newcommand{\eeq}{\end{equation}}

\newcommand{\nn}{\nonumber}

\newcommand{\mbar}{\overline{m}}

\newcommand{\cD}{\mathcal{D}}
\newcommand{\cL}{\mathcal{L}}
\newcommand{\dbar}{\overline{\delta}}

\newcommand{\bd}{\mathbf{d}}
\newcommand{\bdel}{\boldsymbol{\delta}}
\newcommand{\bF}{\mathbf{F}}
\newcommand{\bA}{\mathbf{A}}

\newcommand{\bC}{\mathbf{C}}
\newcommand{\bH}{\mathbf{H}}
\newcommand{\bG}{\mathbf{G}}
\newcommand{\bW}{\mathbf{W}}
\newcommand{\bP}{\mathbf{P}}

\newcommand{\etal}{\emph{et al.}}
\newcommand{\mass}{\mathfrak{m}}

\begin{document}

\preprint{}

 \title{Electromagnetic fields on Kerr spacetime, Hertz potentials and Lorenz gauge}

\author{Sam R. Dolan}
 \email{s.dolan@sheffield.ac.uk}
\affiliation{Consortium for Fundamental Physics,
School of Mathematics and Statistics,
University of Sheffield, Hicks Building, Hounsfield Road, Sheffield S3 7RH, United Kingdom}

\date{\today}

\begin{abstract}
We review two procedures for constructing the vector potential of the electromagnetic field on Kerr spacetime, namely, the classic method of Cohen \& Kegeles, yielding $A^\mu$ in a radiation gauge, and the newer method of Frolov \emph{et al.}, yielding $A^\mu$ in Lorenz gauge. We demonstrate that the vector potentials are related by straightforward gauge transformations, which we give in closed form. We obtain a new result for a separable Hertz potential $H^{\mu \nu}$ such that $A_{\text{lor}}^\mu = \nabla_\nu H^{\mu \nu}$. 
\end{abstract}

\maketitle

%
%

\section{Introduction}
Recent  results from gravitational wave detectors and the Event Horizon Telescope support the hypothesis that the universe is replete with rotating (Kerr) black holes, across a range of mass scales ($\sim 10$---$10^{9} M_{\odot}$). These experimental breakthroughs are underpinned by a solid theoretical understanding of how fields propagate on rotating black hole spacetimes, developed over several decades.

This paper returns once more to the venerable topic of massless (test) fields on rotating black hole spacetimes. This area of enquiry blossomed in the 1970s, after Teukolsky \cite{Teukolsky:1972my,Teukolsky:1973ha,Press:1973zz,Teukolsky:1974yv} showed that certain components of the electromagnetic and gravitational fields on Kerr spacetime satisfy decoupled scalar equations that admit a full separation of variables. Shortly thereafter, Cohen \& Kegeles \cite{Cohen:1974cm,Kegeles:1979an}, Chrzanowski \cite{Chrzanowski:1975wv}, Chandrasekhar \cite{Chandrasekhar:1976,Chandrasekhar:1998}, Wald \cite{Wald:1978vm}, Stewart \cite{Stewart:1978tm}, and others \cite{Guven:1976,Mustafa:1987hertz} developed methods for reconstructing, from scalar potentials, both the vector potential $A^\mu$ of the electromagnetic field, and the metric perturbation $h_{\mu \nu}$ of the gravitational field. Several applications rely on these methods, from the scattering of gravitational waves \cite{Matzner:1978} to gravitational self-force calculations \cite{Whiting:2005hr, Shah:2012gu, Pound:2013faa}.

The classic Hertz potential method \cite{Cohen:1974cm,Chrzanowski:1975wv,Kegeles:1979an} of the 1970s generates fields in a radiation gauge: $l^\mu A_\mu = l^\mu h_{\mu \nu} = 0$, where $l^\mu$ is a principal null direction of the spacetime. However, for certain applications, it is preferable to work with a field in Lorenz gauge\footnote{The gauge condition takes the name of L.~V.~Lorenz (1829--1891) rather than H.~A.~Lorentz (1853--1928).}: $\nabla_\mu A^\mu = 0$ and $\nabla_\mu \overline{h}^{\mu \nu} = 0$, where $\overline{h}^{\mu \nu}$ is the trace-reversed metric perturbation. For example, one might wish to compare with a geometrical-optics approximation, which typically employs this gauge; or evaluate the so-called MiSaTaQuWa self-force formula \cite{Poisson:2011nh}.


By way of motivation, let us consider an electromagnetic field on a general 4D curved spacetime. In the language of forms, the electromagnetic field equations in a region free of charges are
\beq
\bd \bF = 0 , \quad \quad \bdel \bF = 0,  \label{eq:EMfieldeqns}
\eeq
where $\bF$ is the Faraday two-form, $\bd$ denotes the exterior derivative, $\bdel \equiv {}^\star \bd {}^\star$ denotes the coderivative, and ${}^\star$ denotes the Hodge dual operation. By Poincar\'e's lemma, on a contractible domain a form that is closed ($\bd \bF = 0$) is necessarily exact, implying that $\bF = \bd \bA$.\footnote{Similarly, the statement $\bdel \bF = 0$ implies that $\bF = \bdel \bC$ for some three-form $\bC$.} A vector $A^\mu$ corresponding to the one-form $\bA$ is known as the \emph{vector potential}. As is well-known, there is gauge freedom in the vector potential: $\bA$ and $\bA' = \bA + \bd \chi$ (where $\chi$ is an arbitrary scalar field) generate precisely the same Faraday tensor, due to the fundamental identity $\bd\bd = 0$ (from which it also follows that $\bdel \bdel = 0$). A particular gauge that is well-suited to practical calculations is the Lorenz gauge, specified by
\beq
\bdel \bA = 0 \quad \Leftrightarrow \quad \nabla_\mu A^\mu = 0 .
\eeq 
Poincar\'e's lemma applied to the gauge condition then allows one to write $\bA = \bdel \bH$, where $\bH$ is a two-form known as a Hertz potential. 
Once in possession of a Hertz potential, one may generate the vector potential and Faraday tensor by the straightforward application of differential operators.

In 1974, Cohen \& Kegeles \cite{Cohen:1974cm} showed that any spacetime with a shear-free null direction $l^\mu$ (i.e.~any algebraically-special spacetime) admits, in the frequency domain, a Hertz potential that can be constructed from a \emph{scalar} Debye potential \cite{Stewart:1978tm}. Furthermore, on important spacetimes such as Kerr, the decoupled differential equation governing that Debye potential admits a complete separation of variables. This procedure exploits the gauge freedom in $A^\mu$, and the resulting vector potential is \emph{not} in Lorenz gauge; rather, it is in a so-called \emph{radiation gauge} defined by $l^\mu A_\mu = 0$. 

In 2017, Frolov, Krtou\v{s}, Kubiz\v{n}\'ak \& Santos \cite{Frolov:2018ezx} (building on the work of Lunin \cite{Lunin:2017drx}) showed that the Proca equation describing a vector (spin-1) boson of mass $\mass$, that is \cite{Proca:1900nv}
\beq
\bdel \bF + \mass^2 \bA = 0   \label{eq:Proca1}
\eeq
with $\bF = \bd \bA$, admits a complete separation of variables in the frequency domain on Kerr-AdS-NUT spacetimes (a subclass of algebraically-special spacetimes of Petrov type D). In their approach, the potential is $\bA = \mathbf{B} \cdot \boldsymbol{\nabla} Z$, where $\mathbf{B}$ is a certain \emph{polarization tensor} and $Z$ is a scalar potential which admits a separation of variables.
In the case $\mass \neq 0$, it follows from taking the coderivative of Eq.~(\ref{eq:Proca1}) that $\bdel \bA = 0$, and so the vector potential $\bA$ does \emph{not} possess residual gauge freedom; instead, it necessarily satisfies the Lorenz gauge condition. Physically,  the Proca field has three (rather than two) physical polarizations. Taking the massless limit of the Proca equation naturally yields a vector potential for electromagnetism in Lorenz gauge.

This work has three specific aims. First, to review the complementary approaches of Cohen \& Kegeles (1974) and Frolov \etal~(2017) in the context of the 4D Kerr black hole. Second, to identify vector potentials $A_\mu^{\text{(irg/org)}}$ and $A_\mu^{\text{lor}}$, in the ingoing/outgoing radiation gauges and Lorenz gauge respectively, that generate the same Faraday tensor, and to find an explicit gauge transformation between them, that is, a scalar function $\chi$ such that $\bd \chi = \bA^{\text{lor}} - \bA^{\text{(irg/org)}}$. Third, to obtain a Hertz potential $\bH$ which enables one to calculate the Lorenz-gauge potential directly using $\bA^{\text{lor}} = \bdel \bH$. 

In Sec.~\ref{sec:review} we review existing work, covering the Kerr spacetime (\ref{sec:kerr}); Maxwell scalars (\ref{sec:maxwellscalars}); the Teukolsky formalism (\ref{sec:teukolsky}); Hertz potentials for radiation gauge (\ref{sec:hertzradiation}); the Proca equation (\ref{sec:proca}); separation of variables in Lorenz gauge (\ref{sec:lorentz}); and duality (\ref{sec:duality}). The two new results are presented in Sec.~\ref{sec:results}: the aforementioned gauge transformation (\ref{sec:gauge}) and the separable Hertz potential for Lorenz gauge (\ref{sec:hertz}). We conclude with a short discussion (\ref{sec:conclusions}).

\emph{Conventions:} Greek letters $\mu, \nu, \ldots$ are used to denote spacetime indices running from $0$ (the temporal component) to $3$. The covariant derivative of $X_\nu$ is denoted by $\nabla_\mu X_\nu$ or equivalently $X_{\nu;\mu}$, and the partial derivative by $\partial_\mu X_\nu$ or $X_{\nu,\mu}$. The symmetrization (anti-symmetrization) of indices is indicated with round (square) brackets, e.g.~$X_{(\mu\nu)} = \frac{1}{2} (X_{\mu\nu} + X_{\nu\mu})$ and $X_{[\mu\nu]} = \frac{1}{2} (X_{\mu\nu} - X_{\nu\mu})$. For converting between differential forms and tensors we adopt the sign and normalization conventions of Appendix A.2 of Ref.~\cite{Frolov:2017kze}.

\section{Review\label{sec:review}}

 \subsection{The Kerr spacetime and a null tetrad\label{sec:kerr}}
The Kerr spacetime, describing a rotating black hole in vacuum, is characterized by two parameters: mass $M$ and angular momentum $J$, with the latter usually represented by $a \equiv J/M$. The line element describing the (exterior region of) Kerr spacetime in Boyer-Lindquist coordinates $\{t,r,\theta,\phi\}$ is
\begin{align}
ds^2 \equiv g_{\mu \nu} dx^\mu dx^\nu =  -\frac{\Delta_r}{\Sigma} \left( d t-a\sin^2\theta d \varphi \right)^2 + \frac{\Sigma}{\Delta_r} d r^2 + \Sigma\, d\theta^2 + 
 \frac{\sin^2\theta}{\Sigma}\left[(r^2+a^2) d \varphi - a dt\right]^2,\label{eq:linelement}
\end{align}
with~$\Sigma\equiv r^2+a^2\cos^2\theta$, and $\Delta_r \equiv r^2-2Mr+a^2$. 

The inverse metric $g^{\mu \nu}$ can be written in terms of a basis of four null vectors $\{l^\mu, n^\mu, m^\mu, \mbar^\mu\}$ (satisfying $g_{\mu \nu} m^\mu \mbar^\nu = 1 = - g_{\mu \nu} l^\mu n^{\mu} $ with all other scalar products zero) as
\begin{align}
g^{\mu \nu} &= -l^{\mu} n^{\nu} - n^{\mu} l^{\nu} + m^{\mu} \mbar^{\nu} + \mbar^{\mu} m^{\nu}  \\ 
 &= \frac{\Delta_r}{\Sigma} l_{+}^{(\mu} l_{-}^{\nu)} + \frac{1}{\Sigma} m_{+}^{(\mu} m_{-}^{\nu)} .
\end{align}
Here 
\begin{subequations}
\begin{align}
l^\mu &= l_+^\mu, & m^\mu &= \frac{1}{\sqrt{2} (r + i a \cos \theta) } m_+^\mu,   \\
n^\mu &= - \frac{\Delta_r}{2\Sigma} l_-^\mu, & \mbar^\mu &= \frac{1}{\sqrt{2} (r - i a \cos \theta)} m_-^\mu ,
\end{align}
\label{eq:tetrad1}
\end{subequations}
and
\begin{align}
l^\mu_\pm &\equiv \left[ \pm \Delta_r^{-1} (r^2+a^2), 1, 0, \pm \Delta_r^{-1} a \right] , &
m^\mu_{\pm} &\equiv \left[ \pm i a \sin \theta, 0, 1, \pm i \csc \theta \right] .
\label{eq:tetrad2}
\end{align}
The legs $l^\mu$ and $n^\mu$ align with the two principal null directions of the spacetime.

Following standard conventions \cite{Chandrasekhar:1998} we now introduce directional derivatives along the null directions.
The directional derivatives along $\{ l^\mu, n^\mu, m^\mu, \mbar^\mu \}$ are denoted by $\{ D, \Delta, \delta, \dbar \}$, respectively. The directional derivatives along $\{ l_+^\mu, l_-^\mu, m_+^\mu, m_-^\mu \}$ are denoted by
 $\{ \cD, \cD^\dagger, \cL^\dagger, \cL \}$, where 
\begin{subequations}
\begin{align}
\cD \equiv l_+^\mu \partial_\mu &= \partial_r - \frac{i K}{\Delta_r},  & 
\cL^\dagger \equiv m_+^\mu \partial_\mu =  \partial_\theta - Q ,
 \\
\cD^\dagger \equiv l_-^\mu \partial_\mu &= \partial_r + \frac{i K}{\Delta_r} , 
& \cL \equiv m_-^\mu \partial_\mu = \partial_\theta + Q ,
\end{align}
\label{eq:DLoperators}
\end{subequations}
where $K \equiv \omega (r^2 + a^2) - a m$ and $Q \equiv m \csc \theta - a \omega \sin \theta$. 
In addition we define $\cL_n = \cL + n \cot\theta$ and $\cL^\dagger_n = \cL^\dagger + n \cot\theta$. 
Here we assume that these operators act only on quantities with harmonic time dependence $\Psi \equiv e^{-i \omega t + i m \phi}$.

 \subsection{Maxwell scalars\label{sec:maxwellscalars}}
%
%
The six degrees of freedom of a Faraday tensor $F_{\mu \nu}$ are encapsulated in 3 (complex) Maxwell scalars 
\beq
\phi_0 \equiv F_{\mu \nu} l^\mu m^\nu , \quad \phi_2 \equiv F_{\mu \nu} \mbar^\mu n^\nu , \quad \quad \phi_1 \equiv \frac{1}{2} F _{\mu \nu} \left( l^\mu n^\nu - m^\mu \mbar^\nu \right) ,
\label{eq:maxwelldef1}
\eeq
and their 3 complements,
\beq
\phi'_0 \equiv F_{\mu \nu} l^\mu \mbar^\nu , \quad \phi'_2 \equiv F_{\mu \nu} m^\mu n^\nu , \quad \quad \phi'_1 \equiv \frac{1}{2} F _{\mu \nu} \left( l^\mu n^\nu + m^\mu \mbar^\nu \right) .
\label{eq:maxwelldef2}
\eeq
For a real bivector $F_{\mu \nu}$, it follows from the definitions that $\phi'_i = \phi_i^\ast$ ($i=1,2,3$), where $\phi_i^\ast$ is the complex conjugate. For a self-dual bivector it follows rather that $\phi'_i = 0$; and for an anti-self-dual field the converse holds ($\phi_i = 0$). 
For future reference we now introduce four rescaled quantities:
\begin{subequations}
\begin{align}
\Phi_0 &\equiv \phi_0 & \Phi_2 &\equiv 2 (r-ia\cos \theta)^2 \phi_2, \\
\Phi'_0 &\equiv \phi'_0 & \Phi'_2 &\equiv 2 (r+ia\cos \theta)^2 \phi'_2 .
\end{align}
\end{subequations}

 \subsection{The Teukolsky formalism\label{sec:teukolsky}}
 In this section we adopt the Newman-Penrose spin-coefficient formalism \cite{Newman:1961qr}, in which the 24 real connection coefficients for the (rigid) null basis are combined into 12 complex numbers denoted by Greek letters $\{\alpha, \beta, \gamma, \epsilon, \kappa, \rho, \sigma, \mu, \nu, \pi, \tau, \lambda\}$. If the vector $l^\mu$ aligns with a shear-free null geodesic (principal null direction) then $\kappa = \sigma = 0$. If the vector $n^\mu$ also aligns with a principal null direction then $\nu = \lambda = 0$. In a Type-D spacetime, both conditions apply. With these simplifications, Maxwell's equations in vacuum reduce to
 \begin{subequations}
\begin{align}
(\delta^\ast + \pi - 2\alpha) \phi_0 &= (D - 2\rho) \phi_1 , &
(\Delta + \mu - 2 \gamma) \phi_0 &= (\delta - 2 \tau) \phi_1 , \\
(D - \rho + 2 \epsilon) \phi_2 &= (\delta^\ast + 2\pi) \phi_1 , &
(\delta - \tau + 2 \beta) \phi_2 &= (\Delta + 2\mu )\phi_1 . 
\end{align}
\label{eq:maxwelleqsNP}
\end{subequations}
Teukolsky \cite{Teukolsky:1972my,Teukolsky:1973ha} showed that one may then obtain decoupled equations for $\phi_0$ and $\phi_2$ (but not $\phi_1$), viz.,
\begin{subequations}
\begin{align}
\left[ (D - \epsilon + \epsilon^\ast - 2 \rho - \rho^\ast)( \Delta + \mu - 2 \gamma ) - (\delta - \beta - \alpha^\ast - 2 \tau + \pi^\ast)(\delta^\ast + \pi - 2 \alpha) \right] \phi_0 &= 0 , \\
\left[ (\Delta + \gamma - \gamma^\ast + 2\mu + \mu^\ast) (D - \rho + 2\epsilon) - (\delta^\ast + \alpha + \beta^\ast + 2\pi - \tau^\ast)(\delta - \tau + 2\beta) \right] \phi_2 &= 0.
\end{align}
\label{eq:teuk1}
\end{subequations}

\subsubsection{Teukolsky equations}
Upon insertion of the spin coefficients for tetrad (\ref{eq:tetrad1}), namely, $\kappa=\sigma=\nu=\lambda= 0$,
\begin{subequations}
\begin{align}
\rho &= -1/(r-ia\cos\theta) , & \beta &= -\rho^\ast \cot \theta / 2\sqrt{2}, & \pi &= i a \rho^2 \sin \theta / \sqrt{2}, & \alpha &= \pi - \beta^\ast , \\
\tau &= -ia\rho\rho^\ast \sin\theta / \sqrt{2}, & \mu &= \rho^2 \rho^\ast \Delta_r / 2, & \gamma &= \mu + \frac{1}{4}\rho \rho^\ast \Delta' , 
& \epsilon &= 0 , 
\end{align}
\label{eq:npcoeffs}
\end{subequations}
equations (\ref{eq:teuk1}) are \emph{separable} on Kerr spacetime.
With a separable ansatz for the Maxwell scalars, viz.~\cite{Chandrasekhar:1998},
\begin{subequations}
\begin{align}
\phi_0 &= R_{+1}(r) S_{+1}(\theta) \Psi,  \\
2(r-i a \cos \theta)^2 \phi_2 &= R_{-1}(r) S_{-1}(\theta) \Psi , \label{phi2teuk}
\end{align}
\label{eq:RSteuk}
\end{subequations}
one finds that Eqs.~(\ref{eq:teuk1}) yield ordinary differential equations for the functions $R_{\pm1}(r)$ and $S_{\pm1}(\theta)$:
\begin{subequations}
\begin{align}
\left(\Delta_r \cD^\dagger \cD - 2 i \omega r \right) R_{-1} &= \lambda R_{-1} ,   \label{eq:TeukRm} \\
\left(\Delta_r \cD \cD^\dagger + 2 i \omega r \right) \Delta_r R_{+1} &= \lambda \Delta_r R_{+1} ,  \label{eq:TeukRp}  \\
\left(\cL \cL_1^\dagger + 2 a \omega \cos \theta \right) S_{-1} &= -\lambda S_{-1} ,   \label{eq:TeukSm}  \\
\left(\cL^\dagger \cL_1 - 2 a \omega \cos \theta \right) S_{+1} &= -\lambda S_{+1} ,  \label{eq:TeukSp} 
\end{align}
\label{eq:teukolsky-eqns}
\end{subequations}
where $\lambda$ is the separation constant for $s=-1$ \cite{Chandrasekhar:1998}.

 \subsubsection{Teukolsky-Starobinskii identities}
One is \emph{not} free to treat $\phi_0$ and $\phi_2$ as independent variables, even though they satisfy decoupled equations. This is because $\phi_0$ and $\phi_2$ must be mutually consistent with a single Faraday tensor. Instead, one should solve (\ref{eq:TeukRp}) and (\ref{eq:TeukSp}), say, to obtain $\phi_0$ and then deduce $\phi_2$ by solving the first-order equations (\ref{eq:maxwelleqsNP}) consistently. Further analysis of this problem \cite{Starobinskii:1973amp, Teukolsky:1974yv, Chandrasekhar:1976} revealed deep structure which is embodied in the Teukolsky-Starobinsky identities that relate the quantities in Eqs.~(\ref{eq:RSteuk}) in such a way as to a obtain a consistent solution:
\begin{subequations}
\begin{align}
\Delta_r \mathcal{D} \mathcal{D} R_{-1} &= \mathcal{B} \, \Delta_r R_{+1} , \label{TS:Rm} \\
\Delta_r \mathcal{D}^\dagger \mathcal{D}^\dagger \Delta_r R_{+1} &= \mathcal{B} \, R_{-1} , \label{TS:Rp}  \\
\mathcal{L}^\dagger \mathcal{L}_1^\dagger S_{-1} &= \mathcal{B} \, S_{+1} ,  \label{TS:Sm}  \\
\mathcal{L} \mathcal{L}_1 S_{+1} &= \mathcal{B} \, S_{-1} , \label{TS:Sp} 
\end{align}
\end{subequations}
where 
\beq
\mathcal{B} \equiv \sqrt{\lambda^2 + 4 a m \omega - 4 a^2 \omega^2} \label{eq:Bteuk}
\eeq
is the Teukolsky-Starobinsky constant. Having determined $R_{+1}$ and $S_{+1}$ from the differential equations (\ref{eq:TeukRp}) and (\ref{eq:TeukSp}), one may then use (\ref{TS:Rp}) and (\ref{TS:Sp}) to obtain $R_{-1}$ and $S_{-1}$ by the application of differential operators (or vice versa).\footnote{
Here we have used the conventions of Chapter 7 of Chandrasekhar's monograph \cite{Chandrasekhar:1998} in defining $R_{\pm1}$ and $S_{\pm1}$, so that the Teukolsky-Starobinskii identities are as symmetrical as possible. The factor of 2 is included in the definition (\ref{phi2teuk}) to make this consistent. Much of the original literature (e.g.~\cite{Teukolsky:1974yv, Chrzanowski:1975wv}) uses the alternative definitions $\hat{R}_{-1} = -(1/\mathcal{B}) R_{-1}$ and $\hat{R}_{+1} = -2 R_{+1}$. 
} 

With some further work, the scalar $\phi_1$ can be found in terms of $R_{\pm1}$ and $S_{\pm1}$ \cite{Chandrasekhar:1976, Chandrasekhar:1998} and thus (a mode of) the Faraday tensor can be reconstructed in its entirety. However, what is not clear from the results reviewed above is how one could obtain a vector potential $A^\mu$ that generates $F_{\mu\nu}$. A method for this is described in the next section. 


 \subsection{Hertz potentials for radiation gauge\label{sec:hertzradiation}}
In this section, we review the method of Cohen \& Kegeles \cite{Cohen:1974cm,Kegeles:1979an} for obtaining the vector potential in a radiation gauge from a separable Debye potential. The method begins with a modification of the approach outlined in the introduction. Let $\bG$ and $\bW$ denote arbitrary one-forms that we are free to choose; and let $\bA = \bdel \bP - \bG$ where $\bP$ is a Hertz two-form to be determined. Then, in vacuum,
\begin{align}
0 = \bdel \bF &= \bdel \left( \bd \left( \bdel \bP - \bG \right) \right) \nn \\
  &= \bdel \left( \bd \bdel \bP - \bd \bG - {}^\star \bd \bW \right) \nn \\
  &= \bdel \left( \{ \bd \bdel + \bdel \bd \} \bP - \bd \bG - {}^\star \bd \bW \right)
\end{align}
Thus, the field equation $\bdel \bF = 0$ is satisfied by any Hertz potential $\bP$ satisfying
\begin{align}
\hat{\triangle} \bP = \bd \bG + {}^\star \bd \bW  \label{eq:LaplaceBeltrami}
\end{align}
where $\hat{\triangle} \equiv \bd \bdel + \bdel \bd$ is the Laplace-Beltrami (or de Rham) operator. 
This equation has six components, but one may hope to use the gauge freedom in choosing $\bG$ and $\bW$ to seek simplifications. In particular, by choosing an ansatz in which $\bP$ is (anti) self-dual and by choosing $\bG = \pm i \bW$, substantial simplifications occur \cite{Mustafa:1987hertz}. Once $\bP$ is known, it is straightforward to obtain $\bA$ from $\bA = \bdel \bP - \bG$. 


Cohen \& Kegeles \cite{Cohen:1974cm} were the first to show that a self-dual Hertz potential 
\beq
P^{\mu \nu} =  - (l^\mu m^\nu - m^\mu l^\nu) \psi
\eeq
and the choice of gauge terms $G^\mu = -2 \tau \psi l^\mu + 2 \rho \psi m^\mu = i W^\mu$ yields a single decoupled wave equation for the scalar function $\psi$, namely,
\beq
\left[  
(\Delta + \gamma - \gamma^\ast + \mu^\ast) (D + 2 \epsilon + \rho) - (\delta^\ast + \alpha + \beta^\ast - \tau^\ast) (\delta + 2 \beta + \tau) 
\right] \psi = 0 . \label{eq:ck}
\eeq
On Kerr spacetime, Eq.~(\ref{eq:ck}) is the exactly same differential equation as that satisfied by $\Phi_2 = (r-ia\cos\theta)^2 \phi_2$ \cite{Cohen:1974cm, Wald:1978vm}. 

\subsubsection{Ingoing radiation gauge}
A valid solution of Eq.~(\ref{eq:ck}) is $\psi = \psi_{(1)}$ where\footnote{A factor of $1/\mathcal{B}$ is included here for later convenience in comparing to the Lorenz gauge solution.}
\beq
 \psi_{(1)} \equiv \frac{1}{\mathcal{B}} \, R_{-1}(r) S_{-1}(\theta) \Psi .
\eeq 
This yields the vector potential
\begin{align}
A_{(1)}^\mu &= - l^{\mu} (\delta + 2 \beta + \tau) \psi_{(1)} + m^{\mu} (D + \rho) \psi_{(1)} ,  \label{eq:A1}
\end{align}
which generates a Faraday tensor with Maxwell scalars
\begin{subequations}
\begin{align}
\Phi_0 &= 0 & \Phi'_0 &= R_{+1}(r) S_{-1}(\theta) \Psi , \\
\Phi_2 &= 0 & \Phi'_2 &= R_{-1}(r) S_{+1}(\theta) \Psi .
\end{align}
\label{eq:Phirad1}
\end{subequations}
(The middle scalar $\phi_1$ is not zero; it can be found in e.g.~(6.14) of Ref.~\cite{Cohen:1974cm}). 

A complementary solution is generated by the Hertz potential $P^{\mu \nu} =  - (l^\mu \mbar^\nu - \mbar^\mu l^\nu) \psi_{(2)}$ with $\psi_{(2)} = \frac{1}{\mathcal{B}} R_{-1}(r) S_{+1}(\theta) \Psi$, yielding a vector potential
\begin{align}
A_{(2)}^\mu &= - l^{\mu} (\delta^\ast + 2 \beta^\ast + \tau^\ast) \psi_{(2)} + \mbar^{\mu} (D + \rho^\ast) \psi_{(2)} ,  \label{eq:A2}
\end{align}
which generates a Faraday tensor with Maxwell scalars 
\begin{subequations}
\begin{align}
\Phi_0 &= R_{+1}(r) S_{+1}(\theta) \Psi , & \Phi'_0 &= 0 , \\
\Phi_2 &= R_{-1}(r) S_{-1}(\theta) \Psi , & \Phi'_2 &= 0 .
\end{align}
 \label{eq:Phirad2}
\end{subequations}
It straightforward to see the (\ref{eq:A1}) and (\ref{eq:A2}) satisfy the ingoing radiation gauge (IRG) condition, $A_{(1)}^\mu l_\mu = 0 = A_{(2)}^\mu l_\mu$. 

\subsubsection{Outgoing radiation gauge}
A further pair of solutions can be constructed for outgoing radiation gauge (ORG), viz.,
\begin{align}
A_{(3)}^\mu &= \frac{1}{{\rho^\ast}^2} \left[ n^\mu (\delta + \pi^\ast - 2\alpha^\ast) \psi_{(3)} - m^\mu (\Delta + \mu^\ast - 2 \gamma^\ast) \psi_{(3)} \right] , \label{eq:A3}
\end{align}
with $\psi_{(3)} = \frac{2}{\mathcal{B}} R_{+1}(r) S_{-1}(\theta) \Psi$ yielding Maxwell scalars (\ref{eq:Phirad2}); and
\begin{align}
A_{(4)}^\mu &= \frac{1}{(\rho)^2} \left[ n^\mu (\delta^\ast + \pi - 2 \alpha) \psi_{(4)} - \mbar^\mu (\Delta + \mu - 2 \gamma) \psi_{(4)} \right] , \label{eq:A4}
\end{align}
with $\psi_{(4)} = \frac{2}{\mathcal{B}} R_{+1}(r) S_{+1}(\theta) \Psi$ yielding Maxwell scalars (\ref{eq:Phirad1}).
It straightforward to see that (\ref{eq:A3}) and (\ref{eq:A4}) satisfy the ORG condition, $A_{(3)}^\mu n_\mu = 0 = A_{(4)}^\mu n_\mu$. 

For future reference, we now introduce the linear combinations
\begin{align}
A_{\text{(irg)}}^\mu &= A_{(1)}^\mu + A_{(2)}^\mu , &
A_{\text{(org)}}^\mu &= A_{(3)}^\mu + A_{(4)}^\mu , &
A_{\text{(av)}}^\mu &= \frac{1}{2} \left( A_{\text{(irg)}}^\mu + A_{\text{(org)}}^\mu \right). 
\label{eq:irgorg}
\end{align}
The three vector potentials above generate the \emph{same} Faraday tensor; they themselves differ only by the gradient of a scalar.

 \subsection{The Proca equation\label{sec:proca}}
In tensor form, the Proca equation (\ref{eq:Proca1}) is
\beq
\nabla_\nu F^{\mu \nu} + \mass^2 A^\mu = 0, \quad \quad F_{\mu \nu} \equiv \nabla_\mu A_\nu - \nabla_\nu A_\mu. \label{eq:Proca2}
\eeq 
from which it follows as a consequence that $\nabla_\mu A^\mu = 0$ if $\mass \neq 0$.

A separation of variables was recently achieved \cite{Frolov:2018ezx, Lunin:2017drx} by starting with the ansatz
\beq
A^\mu = B^{\mu \nu} \nabla_\nu Z , \label{eq:ansatz}
\eeq 
where $B^{\mu \nu}$ is the \emph{polarization tensor} satisfying \cite{Krtous:2018bvk}
\beq
B^{\mu \nu} \left( g_{\nu \sigma} + i \mu h_{\nu \sigma} \right) = \delta_\sigma^\mu ,  \label{eq:Bdef}
\eeq
$Z$ is a scalar function and $h_{\mu \nu}$ is the closed conformal Killing-Yano tensor (also known as the \emph{principal tensor}). Here $\mu$ is a separation constant to be determined. Following the terminology of Ref.~\cite{Cayuso:2019vyh} we shall call this the \emph{Lunin--Frolov--Krtous--Kubiznak} (LFKK) ansatz.

Solving (\ref{eq:Bdef}) for $B^{\mu \nu}$ yields \cite{Frolov:2018pys, Dolan:2018dqv}
\beq
B^{\mu \nu} = \frac{\Delta_r}{2\Sigma} \left( \frac{l_+^\mu l_-^\nu }{1-i\mu r} + \frac{l_-^\mu l_+^\nu }{1+i\mu r} \right) + \frac{1}{2\Sigma} \left( \frac{m_+^\mu m_-^\nu }{1- a \mu \cos\theta} + \frac{m_-^\mu m_+^\nu }{1+ a\mu \cos\theta} \right) ,
\label{eq:Babalt}
\eeq
Thus the vector potential takes the form
\beq
A^\mu = \frac{1}{2\Sigma} \left[ \frac{\Delta_r \cD^\dag Z}{1 - i \mu r} \, l_+^\mu + \frac{\Delta_r \cD Z}{1 + i \mu r} \, l_-^\mu  +  \frac{\cL Z}{1 - a \mu \cos \theta} \, m_+^\mu +  \frac{\cL^\dagger Z}{1 + a \mu \cos \theta} \, m_-^\mu \right]  .
\label{eq:Alordef}
\eeq
Inserting (\ref{eq:Alordef}) into the Lorenz-gauge condition $\nabla_\mu A^\mu = \frac{1}{\sqrt{-g}} \partial_\mu \left(\sqrt{-g} A^\mu \right) = 0$ leads to
\beq
\cD \left( \frac{\Delta_r \cD^\dagger Z}{1 - i \mu r} \right) +
\cD^\dagger \left( \frac{\Delta_r \cD Z}{1 + i \mu r} \right) + 
\cL^\dagger_1 \left( \frac{\cL Z}{1 - a \mu \cos \theta} \right) +
\cL_1 \left( \frac{\cL^\dagger Z}{1 + a \mu \cos \theta} \right)  = 0 .
\eeq
where operators $\cD$ and $\cL$ are defined in Eq.~(\ref{eq:DLoperators}). 
This equation is clearly separable with the ansatz
\beq
Z = R(r) S(\theta) \Psi,  \quad \quad 
\Psi \equiv e^{-i \omega t} e^{i m \phi} ,
\eeq
leading to
\begin{subequations}
\label{eq:RSsep1}
\begin{align}
\cD \left( \frac{\Delta_r \cD^\dagger R}{1 - i \mu r} \right) + \cD^\dagger \left( \frac{\Delta_r \cD R}{1 + i \mu r} \right) + \kappa_1 R &= 0 ,   \label{eq:Rsep1}  \\
\cL_1^\dagger \left( \frac{\cL S}{1 - a \mu \cos \theta} \right) + \cL_1 \left( \frac{\cL^\dagger S}{1 + a \mu \cos\theta} \right) - \kappa_1 S &= 0 , 
\end{align} 
\end{subequations}
where $\kappa_1$ is a separation constant, to be determined below. 

Employing the ansatz (\ref{eq:ansatz}), Frolov \etal~show that the left-hand side of the field equations (\ref{eq:Proca2}) can be written in the form
\beq
\nabla_\nu F^{\mu\nu} + \mass^2 A^\mu = -B^{\mu \nu} \nabla_\nu J
\eeq
where 
\beq
J = \Box Z - 2 i \xi^\mu A_\mu - \mass^2 Z
\eeq
and $\xi^\mu$ is the time-translation Killing vector. Assuming that $Z$ has harmonic dependence on $t$ and $\phi$ allows one to rewrite equation $J=0$ in the form
\begin{align}
2 \mass^2 \Sigma \, Z &= i \mu r \left( \mathcal{D}^\dagger \left( \frac{\Delta_r \mathcal{D} Z}{1 + i \mu r} \right) - 
\mathcal{D}  \left( \frac{\Delta_r \mathcal{D}^\dagger Z}{1 - i \mu r} \right) \right) 
+ a \mu \cos \theta \left( \cL_1 \left( \frac{\cL^\dagger Z}{1 + a \mu \cos \theta} \right) 
- \cL_1^\dagger \left(\frac{\cL Z}{1 - a \mu \cos \theta} \right)  \right) \nn \\
& \quad - i\mu \left( \frac{\Delta_r \mathcal{D} Z}{1 + i \mu r} - \frac{\Delta_r \mathcal{D}^\dagger Z}{1 - i \mu r} \right)
+ a \mu \sin \theta \left(\frac{\cL^\dagger Z}{1 + a \mu \cos \theta}  - \frac{\cL Z}{1 - a \mu \cos \theta} \right) . \label{eq:J0}
\end{align}
where $\Sigma = r^2 + a^2 \cos^2 \theta$. 
This equation is also separable, and can be written as
\begin{subequations}
\begin{align}
(2 \mass^2 r^2 + \kappa_2) R &= i \mu \left( (r \mathcal{D}^\dagger - 1) \left( \frac{\Delta_r \mathcal{D} R}{1 + i \mu r} \right) 
- (r \mathcal{D} - 1) \left( \frac{\Delta_r \mathcal{D}^\dagger R}{1 - i \mu r} \right) \right) ,  \label{eq:Rsep2} \\
(2 \mass^2 a^2 \cos^2 \theta - \kappa_2) S &= a\mu \left( (\cos \theta \cL_1 + \sin \theta) \left( \frac{\cL^\dagger S}{1 + a \mu \cos \theta} \right) 
- (\cos \theta \cL^\dagger_1 + \sin \theta) \left( \frac{\cL S}{1 - a \mu \cos \theta} \right)  \right) , 
\end{align}
\end{subequations} 
where $\kappa_2$ is a further separation constant. 

Taking the difference between Eq.~(\ref{eq:Rsep1}) multiplied by $\mu^2 r^2$ and Eq.~(\ref{eq:Rsep2}) yields the consistency relation
\beq
\left( \kappa_2 - 2 a \mu (m - a \omega) \right) +  \left(\kappa_1 - 2 \left[ \frac{\omega}{\mu} - \frac{\mass^2}{\mu^2} \right] \right) \mu^2 r^2 = 0 ,  \label{eq:consistency}
\eeq
from which we can read off the values of the separation constants $\kappa_1$ and $\kappa_2$.



With consistency now established, it is straightforward to show from Eqs.~(\ref{eq:RSsep1}) with $\kappa_1 = 2 (\omega/\mu - \mass^2/\mu^2)$ that the Proca equation (\ref{eq:Proca2}) is satisfied if $R(r)$ and $S(\theta)$ obey a pair of second-order ordinary differential equations, viz.,
\begin{subequations}
\begin{eqnarray}
q_r \frac{d}{dr} \left[ \frac{\Delta_r}{q_r} \frac{dR}{dr} \right] + \left[ \frac{K_r^2}{\Delta_r} + \frac{2-q_r}{q_r} \frac{\sigma}{\mu} - \frac{q_r \mass^2}{\mu^2} \right] R(r) &=& 0 ,  \label{eq:Rfkks}  \\
 \frac{q_\theta}{\sin \theta} \frac{d}{d\theta} \left[ \frac{\sin \theta}{q_\theta} \frac{dS}{d\theta} \right] - \left[ \frac{K_\theta^2}{\sin^2 \theta} + \frac{2-q_\theta}{q_\theta} \frac{\sigma}{\mu} - \frac{q_\theta \mass^2}{\mu^2} \right] S(\theta) &=& 0 ,  \label{eq:Sfkks}
\end{eqnarray} 
\label{eq:FKKS}
\end{subequations}
where 
\begin{align}
K_r &= (a^2+r^2)\omega - am, & K_\theta &= m - a\omega \sin^2 \theta, & \Delta_r &= r^2 - 2Mr + a^2,  \nn \\
q_r &= 1+\mu^2 r^2, & q_\theta &= 1 - a^2 \mu^2 \cos^2 \theta, & \sigma &= \omega + a\mu^2 (m-a\omega). \label{eq:sigma}
\end{align}
By direct calculation, the Maxwell scalars (\ref{eq:maxwelldef1}) for the Proca field are \cite{Dolan:2018dqv}
\begin{subequations}
\begin{eqnarray}
\Phi_0 \equiv \phi_0 &=& \phantom{-} \left(\frac{i \mu}{\sqrt{2}} \right) \left( \frac{\cD R}{1+ i \mu r} \right)  \left( \frac{\cL^\dagger S}{1+ a \mu \cos \theta} \right) \Psi,  \label{eq:phi0def} \\
\Phi_2 \equiv 2(r-ia \cos \theta)^2 \phi_2 &=& - \left(\frac{i \mu}{\sqrt{2}} \right) \left( \frac{\Delta_r \cD^\dagger R}{1- i \mu r} \right)  \left( \frac{\cL S}{1- a \mu \cos \theta} \right) \Psi , \label{eq:phi2def} \\
\Phi'_0 \equiv \phi'_0 &=&  \phantom{-} \left(\frac{i \mu}{\sqrt{2}} \right) \left( \frac{\cD R}{1+ i \mu r} \right) \left( \frac{\cL S}{1- a \mu \cos \theta} \right)  \Psi , \\
\Phi'_2 \equiv 2(r+ia \cos \theta)^2 \phi'_2 &=& - \left(\frac{i \mu}{\sqrt{2}} \right) \left( \frac{\Delta_r \cD^\dagger R}{1- i \mu r} \right)  \left( \frac{\cL^\dagger S}{1+ a \mu \cos \theta} \right) \Psi .
\end{eqnarray}
\label{eq:maxwell}
\end{subequations}
The expressions for $\phi_1$ and $\phi'_1$ are somewhat longer and omitted here.

By comparing Eqs.~(\ref{eq:phi0def}--\ref{eq:phi2def}) with Eqs.~(\ref{eq:RSteuk}), we now make the following associations between Teukolsky-like functions $\{ R_{-1}, R_{+1}, S_{-1}, S_{+1} \}$ and the Frolov \etal~functions $\{ R(r), S(\theta) \}$ in the massless limit\footnote{One could, of course, choose to rescale $R_{\pm1} \rightarrow C R_{\pm1}$ and $S_{\pm1} \rightarrow C^{-1} S_{\pm1}$ where $C$ is any constant.} \cite{Dolan:2018dqv}:
\begin{subequations}
\begin{align}
R_{+1} &= \phantom{-} \frac{i \mu}{\sqrt{2}} \frac{\cD R}{(1 + i \mu r)} ,  &
S_{+1} &= \frac{\cL^\dagger S}{1 + a\mu \cos \theta} ,  \\
R_{-1} &= -\frac{i \mu}{\sqrt{2}} \frac{\Delta_r \cD^\dagger R}{(1 - i \mu r)} &
S_{-1} &= \frac{\cL S}{1 - a\mu \cos \theta} .
\end{align}
\label{eq:RSpm}
\end{subequations}
The Maxwell scalars are then simply
\begin{subequations}
\begin{align}
\Phi_0 &= R_{+1} S_{+1} \Psi , & \Phi_2 &= R_{-1} S_{-1} \Psi , \\
\Phi'_0 &= R_{+1} S_{-1} \Psi , & \Phi'_2 &= R_{-1} S_{+1} \Psi . 
\end{align}
\end{subequations}

 \subsection{Electromagnetism in Lorenz gauge\label{sec:lorentz}}
It is straightforward to show, using Eqs.~(\ref{eq:FKKS}) with $\mass^2 = 0$, that the functions defined in (\ref{eq:RSpm}) do indeed satisfy the Teukolsky equations (\ref{eq:teukolsky-eqns}) in the massless limit once we make the identification
\beq
\lambda = -\frac{\omega}{\mu} + (m - a \omega) a \mu , \label{eq:lambda}
\eeq
that is, $\lambda = (\sigma - 2 \omega) / \mu$. 
Solving Eq.~(\ref{eq:lambda}) for the separation constant $\mu$ yields two solutions for each $\lambda$, viz.,
\beq
\mu = \frac{\lambda \pm \mathcal{B}}{2 a (m-a\omega)} = \frac{-2 \omega}{\lambda \mp \mathcal{B}},
\label{eq:mu-massless}
\eeq
where $\mathcal{B}$ is the Teukolsky-Starobinsky constant (\ref{eq:Bteuk}). A further key relationship is that $\sigma / \mu = \pm \mathcal{B}$. We shall denote the solution with the upper sign in Eq.~(\ref{eq:mu-massless}) as $\mu$, and the solution with the lower sign as $\tilde{\mu}$, where 
\beq
\tilde{\mu} = -\frac{1}{a \mu} \frac{\omega}{m - a\omega} .\label{eq:mutilde}
\eeq
We note that $\tilde{\tilde{\mu}} = \mu$, and defer the physical interpretation of this symmetry to the next section.

Eqs.~(\ref{eq:RSpm}) may be inverted to obtain the Frolov \etal~functions $R(r)$ and $S(\theta)$ in terms of the Teukolsky functions:
\begin{subequations}
\begin{align}
\left(\frac{i \mu \mathcal{B}}{\sqrt{2}} \right) R(r)  &= \phantom{-\;\;} (1+i\mu r) \mathcal{D} R_{-1} - i \mu R_{-1} , \\
&= - \left[ (1-i\mu r)  \mathcal{D}^\dagger \Delta_r R_{+1} + i \mu \Delta_r R_{+1} \right] ,
\end{align}
\end{subequations}
and 
\begin{subequations}
\begin{align}
\mathcal{B} \, S(\theta) &= (1 + a \mu \cos \theta)  \left(\mathcal{L}^\dagger + \cot\theta\right) S_{-1} + a \mu \sin \theta S_{-1} , \\
&= (1 - a \mu \cos \theta) \left(\mathcal{L} + \cot\theta \right) S_{+1} - a \mu \sin \theta S_{+1} .
\end{align}
\end{subequations}
From Eqs.~(\ref{eq:RSsep1}), one may derive the following relationships:
\begin{subequations}
\label{eq:RSkeyidentities}
\begin{align}
R(r) &= \frac{i}{\sqrt{2} \, \omega} \left(\cD^\dagger \Delta_r R_{+1} - \cD R_{-1} \right) , \\
S(\theta) &= \frac{\mu}{2 \omega} \left(\cL_1^\dagger S_{-1} + \cL_1 S_{+1} \right) . 
\end{align}
\end{subequations}
These equations will be put to good use in Sec.~\ref{sec:hertz}.

 \subsection{Duality\label{sec:duality}}
It is notable that a single value of the Teukolsky separation parameter $\lambda$ yields two separate values for the LFKK ~separation parameter, $\mu$ and $\tilde{\mu}$, (see Eq.~(\ref{eq:mutilde})) and thus two separate solutions, $\{ R, S \}$ and $\{ \tilde{R}, \tilde{S} \}$. Recent work in Ref.~\cite{Frolov:2018eza} has led to a clear physical interpretation of this observation, which we summarise below.

Let $F_{\mu \nu}$ be the Faraday tensor generated by the vector potential $A^\alpha = B^{\alpha \beta} \nabla_\beta Z$, where $Z= R S \Psi$ and $B^{-1}_{\alpha \beta} = g_{\alpha \beta} + i \mu h_{\alpha \beta}$. The Hodge dual of this Faraday tensor, $\tilde{F}_{\mu\nu} \equiv {}^\star F_{\mu \nu} = \frac{1}{2} \varepsilon_{\mu \nu \alpha \beta} F^{\alpha \beta}$, is generated by the vector potential $\tilde{A}^\alpha = \tilde{B}^{\alpha \beta} \nabla_\beta \tilde{Z}$, where $\tilde{Z} = \tilde{R} \tilde{S} \Psi$ and $\tilde{B}^{-1}_{\alpha \beta} = g_{\alpha \beta} + i \tilde{\mu} h_{\alpha \beta}$, and
\begin{subequations}
\begin{align}
\tilde{R} &=  \frac{\mu}{\omega q_r}  \left(\mu \Delta_r \frac{dR}{dr} + \sigma r \, R \right) ,  \label{eq:tildeR} \\
\tilde{S} &=  \frac{\tilde{\mu}}{\omega q_\theta}  \left( a \mu \sin \theta \frac{dS}{d\theta} + a \sigma \cos \theta \, S \right) . \label{eq:tildeS} 
\end{align}
\end{subequations} 
It is straightforward to verify that $\tilde{R}$ in Eq.~(\ref{eq:tildeR}) satisfies the differential equation (\ref{eq:Rfkks}) with the replacement $\mu \rightarrow \tilde{\mu}$, where $\tilde{\mu}$ is defined in Eq.~(\ref{eq:mutilde}). Similarly, $\tilde{S}$ in Eq.~(\ref{eq:tildeS}) satisfies Eq.~(\ref{eq:Sfkks}) with the same replacement. The dual solution $\tilde{A}^\mu$ is also in Lorenz gauge.

Using these definitions, one may show that $\tilde{R}_{\pm1}$ and $\tilde{S}_{\pm1}$, defined via the dual (`tilded') version of Eqs.~(\ref{eq:RSpm}), are related to $R_{\pm1}$ and $S_{\pm1}$ as follows: 
\begin{align}
\tilde{R}_{\pm1} &= \mp i R_{\pm1} & \tilde{S}_{\pm1} &= \pm S_{\pm1}. \label{eq:RStoRStilde} 
\end{align}
Now we observe, from (\ref{eq:maxwell}) and (\ref{eq:RSpm}), that the Lorenz-gauge vector potential $A^\mu$ generates a Faraday tensor with (normalized) Maxwell scalars 
\begin{align}
\Phi_0 &= R_{+1} S_{+1} \Psi, &\Phi_2 &= R_{-1} S_{-1} \Psi, &\Phi'_0 &= R_{+1} S_{-1} \Psi, &\Phi'_2 &= R_{-1} S_{+1} \Psi. \label{eq:Philorentz}
\end{align}
The dual vector potential $\tilde{A}^\mu$, defined above, generates a Faraday tensor with Maxwell scalars 
\begin{align}
\tilde{\Phi}_0 &= -i \Phi_0, &\tilde{\Phi}_2 &= -i \Phi_2, &\tilde{\Phi}'_0 &= i \Phi'_0, &\tilde{\Phi}'_2 &= i \Phi'_2.  \label{eq:Alormaxwell}
\end{align}
(here we have used the relations (\ref{eq:RStoRStilde})). 
This motivates the introduction of a pair of linear combinations of the original and dual vector potentials, 
\begin{align}
A_\mu^\pm = \frac{1}{2} \left( A_\mu \pm i \tilde{A}_{\mu} \right) .  \label{eq:Aplusminus}
\end{align}
The vector potential $A_\mu^+$ generates a self-dual Faraday tensor with (rescaled) Maxwell scalars
\begin{align}
\Phi_0 &= R_{+1} S_{+1} \Psi, & \Phi_2 &= R_{-1} S_{-1} \Psi, & \Phi'_0 &= \Phi'_2 = 0  , 
\end{align} 
and the vector potential $A_\mu^-$ generates an anti-self-dual Faraday tensor with Maxwell scalars 
\begin{align}
\Phi'_0 &= R_{+1} S_{-1} \Psi, & \Phi'_2 &= R_{-1} S_{+1} \Psi, & \Phi_0 = \Phi_2 = 0 . 
\end{align}

\section{Results\label{sec:results}}

 \subsection{The gauge transformations between radiation and Lorenz gauges\label{sec:gauge}} 
 In this section we find the gauge transformations that translate from ingoing radiation gauge ($l^\mu A^{\text{irg}}_\mu = 0$) and outgoing radiation gauge ($l^\mu A^{\text{org}}_\mu = 0$) to Lorenz gauge ($\nabla_\mu A_{\text{lor}}^\mu = 0$). 
 
 First, we recall that radiation gauge potentials $A^\mu_{(\text{irg/org})}$ were defined in Eqs.~(\ref{eq:irgorg}). 
We see from the sum of Eqs.~(\ref{eq:Phirad1}) and Eqs.~(\ref{eq:Phirad2}) that these solutions have Maxwell scalars which exactly match those in Eq.~(\ref{eq:Philorentz}) for the vector potential in Lorenz gauge. (One can also check the scalars $\phi_1$ and $\phi'_1$). Thus, they generate the same electromagnetic field, and thus there should exist a gauge transformation. 

More explicitly, the vector potential in Lorenz gauge is
\begin{align}
A^\mu_{(\text{lor})} &= \frac{\Psi}{2 \Sigma} \left\{ \frac{\sqrt{2}}{i \mu} \left( - R_{-1} l_+^\mu + \Delta_r R_{+1} l_-^\mu \right) S +  R \left( S_{-1} m_+^\mu + S_{+1} m_-^\mu \right) \right\} \label{eq:Alor2}
\end{align}
where $\{ R_{\pm1}, S_{\pm1} \}$ are Teukolsky functions, $\{ R(r), S(r) \}$ are Frolov \etal~functions, and the two sets are related via Eqs.~(\ref{eq:RSpm}). The expression above was obtained by inserting (\ref{eq:RSpm}) into (\ref{eq:Alordef}).

We first seek a scalar function $\chi^{(\text{irg})}$ such that $\nabla_\mu \chi^{(\text{irg})} = A^{(\text{lor})}_\mu - A^{(\text{irg})}_\mu$. Applying the IRG condition $l^\mu A^{(\text{irg})}_\mu = 0$ yields $\cD \chi^{(\text{irg})} = \l^\mu A^{\text{lor}}_\mu$, that is,
\begin{subequations}
\begin{align}
\cD \chi^{(\text{irg})} &= \frac{\sqrt{2}}{i \mu} R_{+1} S \Psi ,  \\
 &= \frac{\sqrt{2}}{i \mu \mathcal{B}} \cD \cD R_{-1} S \Psi ,
\end{align}
\end{subequations}
where here we have used the Teukolsky-Starobinsky identity (\ref{TS:Rm}). A particular integral of this equation is
\begin{align}
\chi^{(\text{irg})} &= \frac{\sqrt{2}}{i \mu \mathcal{B}} \cD R_{-1} S(\theta) \Psi  , \label{eq:chiirg} \\
 &= -\frac{1}{\mathcal{B}} \cD \left( \frac{\Delta \cD^\dagger R}{1 - i \mu r}  \right) S(\theta) \Psi.  \label{chiirg2}
\end{align}
By inserting expressions (\ref{eq:npcoeffs}) into (\ref{eq:A1}) and (\ref{eq:A2}), it is straightforward but tedious to verify that $\chi^{(\text{irg})}$ in Eq.~(\ref{eq:chiirg}) generates the complete gauge transformation that we seek. 

In a similar way one can find a scalar function $\chi^{(\text{org})}$ such that $\nabla_\mu \chi^{(\text{org})} = A^{(\text{lor})}_\mu - A^{(\text{org})}_\mu$, given by
\begin{subequations}
\begin{align}
\chi^{(\text{org})} &= -\frac{\sqrt{2}}{i \mu \mathcal{B}} \cD^\dagger (\Delta_r R_{+1}) S(\theta) \Psi , \label{chiorg} \\ 
&= -\frac{1}{\mathcal{B}} \cD^\dagger \left( \frac{\Delta \cD R}{1 + i \mu r}  \right) S(\theta) \Psi.  \label{chiorg2}
\end{align}
\end{subequations}

Combining Eq.~(\ref{chiirg2}) and Eq.~(\ref{chiorg2}) yields the gauge transformation for the averaged vector potential  $A^{(\text{av})}_\mu$ defined in Eq.~(\ref{eq:irgorg}). That is, $\nabla_\mu \chi^{(\text{av})} = A^{(\text{lor})}_\mu - A^{(\text{av})}_\mu$ has the solution
\begin{subequations}
\begin{align}
\chi^{(\text{av})} &= \frac{1}{2} \left(\chi^{(\text{irg})} + \chi^{(\text{org})}  \right) , \\
 &= -\frac{1}{2 \mathcal{B}} \left[ \cD \left( \frac{\Delta \cD^\dagger R}{1 - i \mu r}  \right) + \cD^\dagger \left( \frac{\Delta \cD R}{1 + i \mu r}  \right] \right) S(\theta) \Psi , \\
 &= \frac{\omega}{\mu \mathcal{B}} R(r) S(\theta) \Psi .
\end{align}
\end{subequations}
On the final line we made use of Eq.~(\ref{eq:Rsep1}) and (\ref{eq:consistency}). Here we have shown that taking the average of the IRG and ORG solutions does not yield a vector potential in Lorenz gauge; but that the difference is proportional to the gradient of the LFKK function $Z = R(r) S(\theta) \Psi$.  

In a similar way again, one can find the gauge transformation between $A^{(1)}_\mu$ in Eq.~(\ref{eq:A1}) in the IRG and $A^{-}_\mu$ in Eq.~(\ref{eq:Aplusminus}) in the Lorenz gauge; and between $A_\mu^{(2)}$ in Eq.~(\ref{eq:A2}) and $A^{+}_\mu$ in Eq.~(\ref{eq:Aplusminus}).

 \subsection{A separable Hertz potential for Lorenz gauge\label{sec:hertz}}
 
In this section we present a separable Hertz potential $H^{\mu \nu}$ such that $A^\mu_{\text{lor}} = \nabla_{\nu} H^{\mu \nu}$, where $A^\mu_{\text{lor}}$ is the Lorenz gauge vector potential in Eqs.~(\ref{eq:Alordef}) and (\ref{eq:Alor2}). The Hertz potential can be written in separable form as
\begin{align}
H^{\mu \nu} = \frac{i}{2 \sqrt{2} \omega \Sigma} \left( \mathbb{R}^{\mu} \mathbb{S}^{\nu} - \mathbb{S}^{\mu} \mathbb{R}^{\nu} \right) \Psi , \label{eq:Hmunu}
\end{align}
where $\mathbb{R}^\mu(r)$ and $\mathbb{S}^\mu(\theta)$ are vectors defined by
\begin{subequations}
\begin{align}
\mathbb{R}^\mu(r) &=  R_{-1} l_+^\mu - \Delta_r R_{+1} l_-^\mu  , \\
\mathbb{S}^\mu(\theta) &=  S_{-1} m_+^\mu + S_{+1} m_-^\mu  ,
\end{align}
\end{subequations}
Here $R_{\pm1}$ and $S_{\pm1}$ are the Teukolsky functions, $\Psi \equiv \exp(-i \omega t + i m \phi)$ and $\Sigma = r^2 + a^2 \cos^2 \theta$. 
It is straightforward to verify that $\nabla_\nu H^{\mu \nu} = A_{\text{lor}}^\mu$, as follows:
\begin{subequations}
\begin{align}
\nabla_\nu H^{\mu \nu} &= \frac{1}{\sqrt{-g}} \partial_\nu \left( \sqrt{-g} H^{\mu \nu} \right),  \\
 &= \frac{i}{2 \sqrt{2} \omega \Sigma} \frac{1}{\sin \theta}  \partial_\nu \left( \sin \theta \left(   \mathbb{R}^{\mu} \mathbb{S}^{\nu} - \mathbb{S}^{\mu} \mathbb{R}^{\nu} \right) \Psi \right) , \\
 &= \frac{i}{2 \sqrt{2} \omega \Sigma} \left( \frac{1}{\sin \theta} \partial_\nu \left(\sin \theta \, \mathbb{S}^\nu \Psi \right) \mathbb{R}^\mu - \partial_\nu (\mathbb{R}^\nu \Psi) \mathbb{S}^\mu + \mathbb{S}^\nu \tensor{\mathbb{R}}{^\mu _{,\nu}} \Psi - \mathbb{R}^\nu \tensor{\mathbb{S}}{^\mu _{,\nu}} \Psi \right) . \label{eq:Hderiv}
\end{align}
\end{subequations}
The second pair of terms in the parantheses above are zero, as $\mathbb{R}^\mu$ is a function of $r$ only, and $\mathbb{S}^\mu$ is a function of $\theta$ only. Using Eq.~(\ref{eq:RSkeyidentities}), it is straightforward to show that the first pair of terms in Eq.~(\ref{eq:Hderiv}) are
\begin{subequations}
\begin{align}
\partial_\nu (\tensor{\mathbb{R}}{^\nu} \Psi) &= (\cD R_{-1} - \cD^\dagger \Delta_r R_{+1}) \Psi , \nn \\
&= \sqrt{2} i \omega R(r) \Psi ,  \\
\frac{1}{\sin \theta} \partial_\nu \left( \sin \theta \, \tensor{\mathbb{S}}{^\nu} \Psi \right) &= \left( \cL_1^\dagger S_{-1} +  \cL_1 S_{+1} \right) \Psi , \nn \\ 
&= \frac{2 \omega}{\mu} S(\theta) \Psi
\end{align}
\end{subequations}
where $R(r)$ and $S(\theta)$ are the functions of Frolov \etal \; Inserting these expressions into (\ref{eq:Hderiv}) yields
\begin{align}
\nabla_\nu H^{\mu \nu} &= \frac{\Psi}{\sqrt{2} i \mu \Sigma} \left( - R_{-1} l_+^\mu + \Delta_r R_{+1} l_-^\mu \right) S + \frac{1}{2 \Sigma} R \left( S_{-1} m_+^\mu + S_{+1} m_-^\mu \right) ,
\end{align}
which matches Eq.~(\ref{eq:Alor2}).

This Hertz potential is not unique, as we can add any divergence-free bivector to it without affecting the defining relationship $A^\mu_{\text{lor}} = \nabla_{\nu} H^{\mu \nu}$. In particular, one can add scalar multiples of the Faraday tensor $F_{\mu \nu}$ itself.

A second linearly-independent solution, also in Lorenz gauge, is obtained by inserting 
\begin{subequations}
\begin{align}
\mathbb{R}^\mu(r) &=  R_{-1} l_+^\mu + \Delta_r R_{+1} l_-^\mu  , \\
\mathbb{S}^\mu(\theta) &=  S_{-1} m_+^\mu - S_{+1} m_-^\mu  ,
\end{align}
\end{subequations}
into Eq.~(\ref{eq:Hmunu}).

\section{Discussion and conclusion\label{sec:conclusions}}
We have shown that the recent method of Frolov \etal~\cite{Frolov:2018eza,Frolov:2018ezx,Frolov:2018pys} and Lunin \cite{Lunin:2017drx} (in the massless case $\mass = 0$) is closely related to the classic method of Cohen \& Kegeles \cite{Cohen:1974cm} for constructing a vector potential on Kerr spacetime. More precisely, we have identified the gauge transformation between the Lorenz-gauge vector potential ($\nabla_\mu A_{\text{lor}}^\mu = 0$) of the former and the radiation-gauge vector potentials ($A_{(\text{irg})}^\mu l_\mu = 0$ and $A_{(\text{org})}^\mu n_\mu = 0$) of the latter. We have found a Hertz potential (\ref{eq:Hmunu}) that generates the Lorenz-gauge vector potential on Kerr spacetime. This Hertz potential has a neat separable form: $\Sigma H^{\mu \nu} \Psi^{-1}$ is the exterior product of $\mathbb{R}^\mu(r)$, a vector constructed from the two principal null directions of the Type-D spacetime which is a function of $r$ only, and $\mathbb{S}^\mu(\theta)$, a vector constructed from directions in the orthogonal two-space which is a function of $\theta$ only.

In 1976, Chandrasekhar \cite{Chandrasekhar:1976} described a method for constructing the vector potential in an arbitrary gauge. This method did not, however, lead to the identification of a vector potential in Lorenz gauge. In retrospect, it seems plausible that Lorenz gauge did not emerge naturally at that time because a separation constant was not introduced in Eq.~(36) of that work. 

Curiously, the Lorenz-gauge solution described here satisfies a rather specific constraint,
\beq
h^{\mu \nu} F_{\mu \nu} = -2 \xi^\mu A_\mu, 
\eeq
where $h_{\mu \nu}$ is the conformal Killing-Yano tensor, and $\xi^\mu = \frac{1}{3} \nabla_{\nu} h^{\nu \mu} =  [1,0,0,0]$ is the time-translation Killing vector field. To obtain a more general Lorenz-gauge solution one may make a (restricted) gauge transformation $A_{\mu} \rightarrow A_{\mu} + \nabla_\mu \chi$ where $\chi$ is any scalar field satisfying $\Box \chi = 0$.  

An open question is whether the Lorenz-gauge methods of Frolov \etal~\cite{Frolov:2018ezx,Frolov:2018pys} can be extended to obtain metric perturbations $h_{\mu \nu}$ on the Kerr spacetime. For non-spherically symmetric black holes, there are presently no simple, uncoupled equations to obtain metric perturbations themselves $h_{\mu \nu}$ \cite{Whiting:2005hr}. Instead, one may reconstruct the metric perturbation in a radiation gauge $h_{\mu \nu} l^\nu = 0$ by following the method of Chrzanowski \cite{Chrzanowski:1975wv} which extends the approach of Cohen \& Kegeles to the spin-2 sector (see also Refs.~\cite{Guven:1976,Stewart:1978tm,Wald:1978vm,Kegeles:1979an}). For certain applications, such as gravitational self-force calculations, it would be advantageous to construct the metric perturbation in Lorenz gauge, or another regular gauge. In the presence of sources, the metric perturbation constructed in radiation gauge is known to have spurious string-like gauge singularities \cite{Barack:2001ph, Pound:2013faa}. This is a possible concern for second-order calculations \cite{Barack:2018yvs}, in which the first-order metric perturbation acts as a source for the second-order field.

It is notable that the Lorenz-gauge vector potentials on a Ricci-flat spacetime, which satisfy $\Box A^\mu = 0$ and $\nabla_\mu A^\mu = 0$, can be used to generate pure-gauge metric perturbations $h_{\mu \nu} = 2 A_{(\mu ; \nu)}$ that are tracefree $\tensor{h}{^\mu _\mu} = 0$ and in Lorenz gauge $\nabla_\mu \overline{h}^{\mu \nu} = 0$. This, or the separable form of the Hertz potential in Eq.~(\ref{eq:Hmunu}), may give some guide to the form of the ansatz to use in extending to the gravitational sector.

\begin{acknowledgments}
With thanks to Jake Shipley, Marc Casals and Marco Cariglia for discussions, and to Germain Rousseaux and Jos\'e Lemos for email correspondence. I acknowledge financial support from the European Union's Horizon 2020 research and innovation programme under the H2020-MSCA-RISE-2017 Grant No.~FunFiCO-777740, and from the Science and Technology Facilities Council (STFC) under Grant No.~ST/P000800/1.
\end{acknowledgments}

\appendix
 \section{Forms and tensors\label{forms}}
A $p$-form $\boldsymbol{\alpha}$ is equivalent to a completely antisymmetric tensor $\alpha_{\mu_1 \ldots \mu_p}$ of rank $(0,p)$. 
The Hodge dual of a $p$-form $\boldsymbol{\alpha}$ is a $(4-D)$-form ${}^\star \boldsymbol{\alpha}$ defined by
\beq
({}^\star \alpha)_{\mu_{p+1} \ldots \mu_{D}} = \frac{1}{p!} \alpha^{\nu_1 \ldots \nu_{p}} \varepsilon_{\nu_1 \ldots \nu_{p} \mu_{p+1} \ldots \mu_{D}} .
\eeq
where $\varepsilon_{\mu_1 \ldots \mu_D}$ is the Levi-Civita tensor. 
A key property of the Hodge dual of a two-form $\bF$ in $D=4$ Lorenzian spacetimes is that ${}^\star {}^\star \bF = - \bF$, leading to a natural role for complex numbers. From an arbitrary (real or complex) bivector $F_{\mu \nu}$, one may construct a self-dual version $\mathcal{F}_{\mu \nu} = F_{\mu \nu} - i {}^\star F_{\mu \nu} $ satisfying $({}^\star \mathcal{F})_{\mu \nu} =  i \mathcal{F}_{\mu \nu}$. The complex bivectors $l \wedge m$, $\mbar \wedge n$ and $\frac{1}{2} (l \wedge n - m \wedge \mbar)$ span the space of self-dual bivectors, where $\{l, n, m, \mbar\}$ is any complex null tetrad. Here $\wedge$ denotes the exterior product, such that $(l\wedge m)^{\mu \nu} = l^\mu m^\nu - m^\mu l^\nu$, etc. The exterior derivative $\bd$ acts on a $p$-form to produce a $(p+1)$-form, and the coderivative $\bdel$ acts on a $p$-form to yield a $(p-1)$-form, according to the rules
\begin{subequations}
\begin{align}
(d \alpha)_{\mu_0 \ldots \mu_p} &= (p + 1) \nabla_{[\mu_0} \alpha_{\mu_1 \ldots \mu_p]} , \\
(\delta \alpha)_{\mu_2 \ldots \mu_p} &= - \nabla^{\mu_1} \alpha_{\mu_1 \ldots \mu_p} . 
\end{align}
\end{subequations}
Further details are given in e.g.~Appendix A.2 of Ref.~\cite{Frolov:2017kze}.

\bibliographystyle{apsrev4-1}
\bibliography{refs}

\end{document}